\documentclass[conference]{IEEEtran}
\IEEEoverridecommandlockouts
\usepackage[table,xcdraw]{xcolor}
\usepackage{cite}
\usepackage{amsmath,amssymb,amsfonts}
\usepackage{algorithmic}
\usepackage{graphicx}
\usepackage{textcomp}
\usepackage{xcolor}
\usepackage{textcomp}
\usepackage{svg}
\usepackage{pgf}
\usepackage{amssymb}
\usepackage{multirow}
\usepackage{tabularx}

\usepackage{listings}
\def\BibTeX{{\rm B\kern-.05em{\sc i\kern-.025em b}\kern-.08em
    T\kern-.1667em\lower.7ex\hbox{E}\kern-.125emX}}
\begin{document}

\newtheorem{definition}{Definition}

\newcommand{\pandas}{\texttt{Pandas}}
\newcommand{\cylon}{\textit{Cylon}}
\newcommand{\cflow}{\textit{CylonFlow}}
\newcommand{\pycylon}{\textit{PyCylon}}
\newcommand{\gcylon}{\textit{GCylon}}

\newcommand{\todo}[1]{{\color{red} TODO: #1}}


\title{Supercharging Distributed Computing Environments For High Performance Data Engineering}

\author{
\IEEEauthorblockN{Niranda Perera\IEEEauthorrefmark{1}, 
Kaiying Shan\IEEEauthorrefmark{2}, 
Supun Kamburugamuve\IEEEauthorrefmark{3},
Thejaka Amila Kanewela\IEEEauthorrefmark{3},
Chathura Widanage\IEEEauthorrefmark{3},\\
Arup Kumar Sarker\IEEEauthorrefmark{2},
Mills Staylor \IEEEauthorrefmark{2},
Tianle Zhong\IEEEauthorrefmark{2}, 
Vibhatha Abeykoon\IEEEauthorrefmark{3},
Geoffrey Fox\IEEEauthorrefmark{2}\IEEEauthorrefmark{4}
}

\IEEEauthorblockA{\IEEEauthorrefmark{1}Luddy School of Informatics, Computing, and Engineering, Indiana University,\\
Bloomington, IN 47408, USA\\
dnperera@iu.edu}
\IEEEauthorblockA{
\IEEEauthorrefmark{2}University of Virginia, Charlottesville, VA 22904, USA
\{ks5qug, djy8hg, qad5gv, fad3ew\}@virginia.edu}
\IEEEauthorblockA{\IEEEauthorrefmark{3}Indiana University Alumni, Bloomington, IN 47405, USA\\
supun@apache.org, \{thejaka.amila, chathurawidanage\}@gmail.com}
\IEEEauthorblockA{\IEEEauthorrefmark{4}Biocomplexity Institute and Initiative, University of Virginia, Charlottesville, VA\\
22911, USA\\
vxj6mb@virginia.edu}
}

\maketitle

\begin{abstract}
The data engineering and data science community has embraced the idea of using Python \& R dataframes for regular applications. Driven by the big data revolution and artificial intelligence, these applications are now essential in order to process terabytes of data. They can easily exceed the capabilities of a single machine, but also demand significant developer time \& effort. Therefore it is essential to design scalable dataframe solutions. There have been multiple attempts to tackle this problem, the most notable being the dataframe systems developed using distributed computing environments such as Dask and Ray. Even though Dask/Ray distributed computing features look very promising, we perceive that the Dask Dataframes/Ray Datasets still have room for optimization. In this paper, we present \cflow{}, an alternative distributed dataframe execution methodology that enables state-of-the-art performance and scalability on the same Dask/Ray infrastructure (thereby \textit{supercharging} them!). To achieve this, we integrate a \textit{high performance dataframe} system \cylon{}, which was originally based on an entirely different execution paradigm, into Dask and Ray. Our experiments show that on a pipeline of dataframe operators, \cflow{} achieves $30\times$ more distributed performance than Dask Dataframes. Interestingly, it also enables superior sequential performance due to the native C++ execution of \cylon{}. We believe the success of \cylon{} \& \cflow{} extends beyond the data engineering domain, and can be used to consolidate high performance computing and distributed computing ecosystems.

\end{abstract}



\section{Introduction}\label{sec:intro}

The data engineering domain has expanded at a staggering pace over the past few decades, predominantly owing to the emergence of the \textit{Big Data revolution}, machine learning (ML), and artificial intelligence (AI). In today's information age, data is no longer referred to in megabytes, files or spreadsheets, but in giga/terabytes and object stores. This overabundance of data takes up a significant amount of developer time for data preprocessing when it would be better served focusing their attention on building data engineering models. Therefore, it is crucial to improve the performance of these data preprocessing stages in order to build efficient data engineering pipelines. Data preprocessing has been traditionally done on database systems using a structured query language (SQL), but more recently Python and R programming languages have taken over these SQL workloads. Functional interface, interactive programming environment, and interpreted execution of these languages provide a more user-friendly developing ecosystem for modern-day engineers. 

The Python library \pandas{} has been at the forefront of this transformation, and has played a vital role in popularizing Python for data exploration. In this paper, we focus mainly on the \textit{Dataframe (DF)} API, which is at the heart of the \pandas{} ecosystem. The concept of a DF is not unique to \pandas{}; in fact, it originated from S language in the 1990s, and was subsequently popularized by the R language. However \pandas{} dominates the field with over 100 million monthly downloads consistently, according to the \textit{PyPI} package index stats\cite{PyPIDown82:online}. Despite this popularity, both \pandas{} \& R DF run into performance limitations even on moderately large datasets \cite{widanage2020high,perera2022high,petersohn2020towards}.  For example, in an Intel\textsuperscript{\textregistered} Xeon\textsuperscript{\textregistered} Platinum 8160 high-end workstation with 240GB memory, it takes around 700s to \texttt{join} two DFs with 1 billion rows each for \texttt{pandas}, whereas traversing each dataframe only takes about 4s. On the other hand, today's computer hardware carries plenty of computing power with a large amount of memory. On-demand elastic cloud computing services enable work to be done on thousands of such nodes with the touch of a button. As such, there are plenty of resources at our disposal to develop more efficient distributed data engineering solutions. 

Hadoop YARN, Dask, and Ray are just a few distributed execution runtimes capable of managing thousands of computing resources under their purview. These engines were predominantly developed by the distributed and cloud computing communities, and provide application program interfaces (API) to conveniently submit user logic across many nodes. They employ several execution models such as asynchronous many-tasks (AMT), actors, etc. In the data engineering community, we have seen several frameworks attempting to leverage these distributed runtimes to develop distributed dataframe (DDF) solutions. Spark SQL RDDs \& Datasets was a breakthrough framework on this front, significantly improving the traditional map-reduce paradigm \cite{zaharia2012resilient}. Dask developed its own take on DDFs, Dask DDF, closely followed by Ray with Ray-Datasets. Modin is the latest attempt to develop scalable DF systems \cite{petersohn2020towards}, which is also built on top of Dask \& Ray. However in practice, we have encountered several performance limitations with these systems \cite{widanage2020high,perera2022high}, as discussed in Section \ref{sec:exp}. 

Traditionally, the high performance computing (HPC) community has been developing solutions based on the bulk synchronous parallel (BSP) execution model using the message passing interface (MPI) specification. They have been able to achieve commendable scalability \& performance on thousands of CPU cores (and on supercomputers). In a previous publication we developed an alternative to the existing DDFs named \cylon{} \cite{widanage2020high}, which looks at the problem from the HPC point of view. \cylon{} employs BSP model for DDF operator execution, and works on top of MPI runtimes (OpenMPI, MPICH, IBM Spectrum MPI, etc). Due to superior scalability and HPC descent, we differentiate \cylon{} as a \textit{high performance DDF (HP-DDF)} implementation. Apart from running on BSP, another notable feature in HP-DDFs is the use of an optimized communication library. 

\begin{figure}[hptb]
\centering
\includegraphics[width=\linewidth]{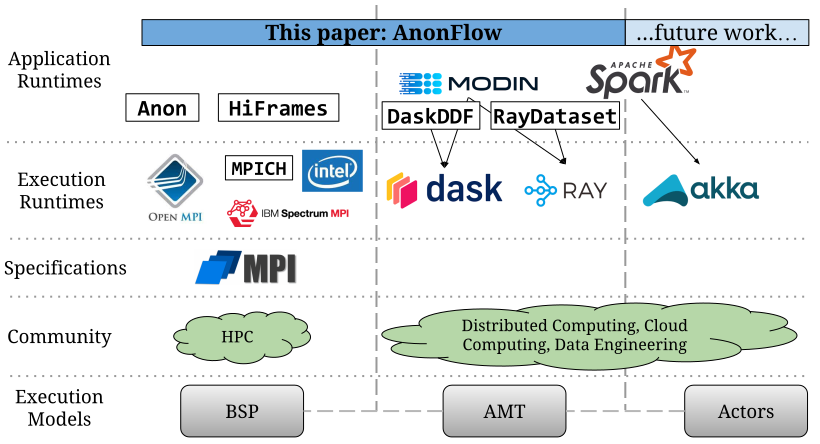}
\caption{Current Status Quo \& \cflow{} Contribution}
\label{fig:cflow}
\end{figure}

Even though \cylon{} has been able to achieve above-par scalability compared to the popular DDF systems, it is tightly coupled to the MPI ecosystem. As we discuss in Section \ref{sec:hp_ddf}, this limits us from extending HP-DDF concept to distributed computing environments such as Dask or Ray. In this paper, we propose an alternative execution methodology to resolve this limitation. Our objective is to integrate \cylon{} with other execution runtimes without compromising its scalability and performance. It is a bipartite solution: 1. creating a stateful pseudo-BSP environment within the execution runtime resources; 2. using a modularized \textit{communicator} that enables plugging-in optimized communication libraries. We named it \cflow{} because the idea carries parallels to workflow management. We demonstrate the robustness of this idea by implementing \cylon{} HP-DDF runtimes on top of Dask (\cflow{}-on-Dask) and Ray (\cflow{}-on-Ray) that outperform their own DDF implementations. We also confirm that the idea gives comparable or better results than MPI-based \cylon{} DDF on the same hardware. With \cflow{}, we have now enabled HP-DDFs from anywhere to personal laptops or exascale supercomputers. As depicted in Figure \ref{fig:cflow}, it consolidates disparate execution models and communities under a single application runtime. To the best of our knowledge, this is the first attempt to adapt high performance data engineering constructs to distributed computing environments. We believe that the methodology behind \cflow{} extends beyond the data engineering domain, and it could be used to execute many HPC applications on distributed computing environments.

\section{Distributed Computing Models \& Libraries}\label{sec:models}

In order to understand the design and implementation of both \cylon{} \& \cflow{}, it is important to discuss the existing distributed computing models and prevalent libraries that implement them. A distributed computing model provides an abstract view of how a particular problem can be decomposed and executed from the perspective of a machine. It describes how a distributed application expresses and manages parallelism. \textit{Data parallelism} executes the same computation on different parts (partitions) of data using many compute units. We see this at the instruction level, \textit{single-instruction multiple-data (SIMD)}, as well as in program level \textit{single-program multiple-data (SPMD)}. On the other hand, \textit{task parallelism} involves executing multiple tasks in parallel over many compute units. This is a form of \textit{multiple-program multiple-data (MPMD)} at the program level.  

    \subsection{Bulk Synchronous Parallel (BSP)}
    
    \textit{BSP} or Communicating Sequential Processors (CSP) model \cite{valiant1990bridging, fox1989solving} is the most common model that employs SPMD \& \textit{data parallelism} over many compute nodes. Message Passing Interface (MPI) is a formal specification of BSP model that has matured over 30+ years. OpenMPI, MPICH, MSMPI, IBM Spectrum MPI, etc. are some notable implementations of this specification. MPI applications display \textit{static parallelism} since most often parallelism needs to be declared at the initiation of the program. From the point of view of the data, this would mean that the data partitions are tightly coupled to the parallelism. At the beginning of the application, data partitions would be allocated to executors/workers. Executors then own data partitions until the end of the application and perform computations on them. When the workers reach a communication operation in the program, they synchronize with each other by passing messages. Many high performance computing (HPC) applications use the BSP model on supercomputing clusters and have shown admirable scalability. However, only a handful of data engineering frameworks have adopted this model, including \textit{Twister2} \cite{kamburugamuve2020twister2} \& \cylon{}. 
    
    \subsection{Asynchronous Many-Tasks (AMT)}
    
    \textit{AMT} model relaxes the limitations of BSP by decomposing applications into independent transferable sub-programs (many tasks) with associated inputs (data dependencies). AMT runtimes usually manage a distributed queue that accepts these tasks (\textit{Manager/Scheduler}). A separate group of executors/workers would execute tasks from this queue, thus following MPMD \& \textit{task parallelism}. Dependencies between tasks are handled by the scheduling order. This allows the application to set parallelism on-the-fly, and the workers are allowed to scale up or down, leading to \textit{dynamic parallelism}. AMT also enables better resource utilization in multi-tenant/multi-application environments by allowing free workers to pick independent tasks, thereby improving the overall throughput of the system. Furthermore, task parallelism enables task-level fault tolerance where failed tasks can be rerun conveniently. These benefits may have prompted many distributed dataframe runtimes, including Dask DDF \& Ray Datasets, to choose AMT as the preferred execution model.
    
    \subsection{Actors}
    
    \textit{Actor} model was popularized by \textit{Erlang}\cite{armstrong2010erlang}. An actor is a primitive computation which can receive messages from other actors, upon which they can execute a computation, create more actors, send more messages, and determine how to respond to the next message received. Compared to executors and tasks in AMT, actors manage/maintain their own state, and the state may change based on the computation/communication. Messages are sent asynchronously and placed in a \textit{mailbox} until the designated actor consumes them. Akka is a popular actor framework which was used as the foundation for the Apache Spark project. Interestingly, Dask and Ray projects also provide an actor abstraction on top of their distributed execution runtimes mainly aimed at reducing expensive state initializations.

\section{Distributed Data Dataframes (DDF)}\label{sec:ddf}

With the exponential growth in dataset sizes, it is fair to conclude that data engineering applications have already exceeded the capabilities of a single workstation node. Modern hardware offers many CPU cores/threads for computation, and the latest cloud infrastructure enables users to spin many such nodes instantaneously. As a result, there is abundant computing power available at users' disposal, and it is essential that data engineering software make use of it. Furthermore, every AI/ML application requires a pre-processed dataset, and it is no secret that data pre-processing takes significant developer time and effort. Several AI/ML surveys suggest that it could even be more than 60\% of total developer time \cite{anaconda-survey}. For these reasons, using scalable \textit{distributed dataframe (DDF)} runtime could potentially improve the efficiency of data engineering pipelines immensely. Based on our experiments with some widely used DDF systems (Section \ref{sec:exp}), we believe that the idea of a \textit{high performance scalable DDF runtime} is still a work in progress. 

    \subsection{Dataframes (DF)}
    
    Let us first define a dataframe. We borrow definitions from the relations terminology proposed by Abiteboul et al \cite{abiteboul1995foundations}. Similar to SQL tables, DFs contain heterogeneously typed data. These elements originate from a known set of \textit{domain}s, $Dom = \{dom_1, dom_2, ...\}$. For a DF, these \textit{domain}s represent all the data types it supports. A \textbf{Schema} of a DF $S_M$ is a tuple $(D_M, C_M)$, where $D_M$ is a vector of $M$ domains and $C_M$ is a vector of $M$ corresponding column labels. Column labels usually belong to \textit{String}/\textit{Object} domain. A \textbf{Dataframe (DF)} is a tuple $(S_M, A_{NM}, R_N)$, where $S_M$ is the Schema with $M$ domains, $A_{NM}$ is a 2-D array of entries where actual data is stored, and $R_N$ is a vector of $N$ row labels belonging to some domain. \emph{Length} of the dataframe is $N$, i.e. the number of rows.

    Heterogeneously typed schema clearly distinguishes DFs from multidimensional arrays or tensors. However data along a column is still homogeneous, so many frameworks have adopted a columnar data format which enables vectorized computations on columns. A collection of \texttt{numpy NDArrays} would be the simplest form of DF representation. Alternatively, Apache Arrow columnar format\cite{apacheArrowColumnar} is commonly used by many DF runtimes. Arrow arrays are composed of multiple buffers such as data, validity and offsets for variable length types (e.g. \texttt{string}).

    As identified in previous literature, many commonly used DF operators are defined over the vertical axis (row-wise)\cite{petersohn2020towards, perera2022high}. Even though columnar representation allows contiguous access along a column, it makes indexing or slicing rows non-trivial. Furthermore, many DF operators are defined on a set of \textit{key columns}, while the rest (i.e. \textit{value columns}) move along with the keys. As a consequence, traditional BLAS (basic linear algebra subprograms) routines cannot be directly used for DF operators.

    \subsection{DDF System Design}
    
    The composition of a DF introduces several engineering challenges in designing distributed DF systems. Similar to any distributed/parallel system design, let us first examine the computation and communication aspects broadly.
    
        \subsubsection{Computation}
        
        \begin{figure}[hptb]
        \begin{tabular}{c}
        \includegraphics[width=\linewidth]{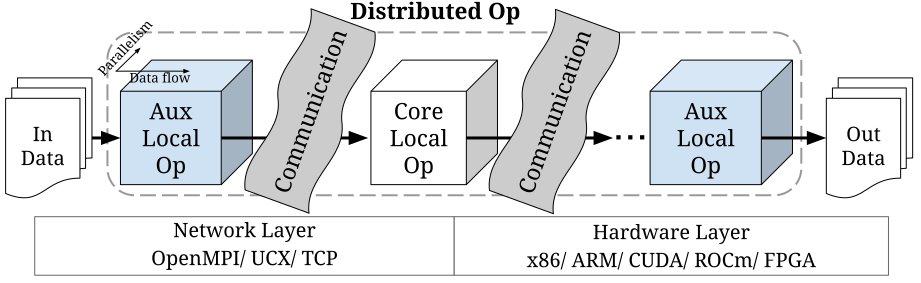}\\
        \includegraphics[width=0.8\linewidth]{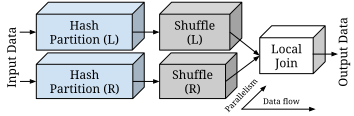}
        \end{tabular} 
        \caption{Distributed DDF Sub-operator Composition \cite{perera2022high} (Bottom: Join Operator Example)}
        \label{fig:dist-join}
        \end{figure}
        
\begin{figure}[hptb]
        \begin{center}
        \includegraphics[width=0.8\linewidth]{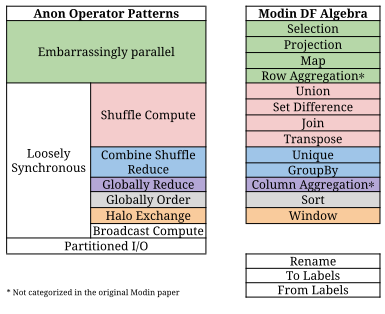}
        \end{center}
        \caption{\cylon{} Operator Patterns \& Modin DF Algebra}
        \label{fig:cylon_modin}
        \end{figure}

        Petersohn et al \cite{petersohn2020towards} recognize that many \pandas\ operators can potentially be implemented by a set of core operators, thereby reducing the burden of implementing a massive DDF API. Correspondingly, in a recent publication we observed that DF operators follow several generic distribution execution patterns \cite{perera2022high}. The \textit{pattern} governs how these sub-operators are arranged in a directed acyclic graph (DAG). We also identified that a DDF operator consists of three major sub-operators: 1. core local operator; 2. auxiliary local operators; and 3. communication operators. Figure \ref{fig:dist-join} depicts a distributed \texttt{join} operation composition, and Figure \ref{fig:cylon_modin} shows the relationship between the concepts of \cylon{} and Modin. A framework may choose to create tasks (i.e. the definition for a unit of work) for each of these sub-operators. A typical application would be a pipeline of multiple DDF operators. 

        When using the AMT model, these tasks would be further expanded for each data partition (parallelism). Every task would produce input data for subsequent tasks. This dataflow governs the dependencies between tasks. When there are several operators in a DAG, it is common to see multiple local tasks grouped together. An \textit{execution plan optimizer} may identify such tasks and coalesce them together into a single local task. We see these optimizations in the Apache Spark SQL \textit{Tungsten} optimizer \cite{Catalyst12:online}. 
        Previously mentioned in Section \ref{sec:intro}, data parallelism is natively supported by the BSP model. Since the executors own data partitions until the end of an application, they have the ability to perform all local compute tasks until they reach a communication boundary. As such, coalescing subsequent local tasks are inherently supported by the model itself compared to AMT.

        \subsubsection{Communication}
        
        Implementing DDF operators requires point-to-point (P2P) communication, as well as complex message passing between worker processes. We have identified several such collective communication routines, such as shuffle (all-to-all), scatter, (all)gather, broadcast, (all)reduce, etc, that are essential for DDF operators \cite{perera2022high}. Typically, communication routines are performed on data buffers (ex: MPI, TCP), but the DF composition dictates that these routines be extended on data structures such as DFs, arrays, and scalars. Such data structures may be composed of multiple buffers (Section \ref{sec:intro}) which could further complicate the implementation. For example, \texttt{join} requires a DF to be \texttt{shuffled}, and to do this we must \texttt{AllToAll} the buffer sizes of all columns (counts). We then shuffle column data based on these counts. 
        In most DF applications, communication operations may take up significant wall time, creating critical bottlenecks. This is evident from Section \ref{sec:comm_comp}, where we evaluate the distribution of communication and computation time over several DF operator patterns.  Moreover, developer documentation of Spark SQL, Dask DDF, Ray Datasets, etc, provide special guidelines to reduce \texttt{shuffle} routine overheads \cite{daskshuffle:online,ray:online}.

        While these communication routines can be implemented ingenuously using point-to-point message passing, implementation of specialized algorithms has shown significant performance improvements \cite{bruck1997efficient, thakur2005optimization, traff2014implementing}. For instance, OpenMPI implements several such algorithms for its collective communications, which can be chosen based on the application. Typically in AMT run-times, communications between tasks are initiated with the help of a \textit{Scheduler}. 
        Another approach is to use a distributed object store or a network file system to share data rather than sending/receiving data explicitly, although this could lead to severe communication overhead. 
        



    
    \subsection{DDF Systems Examined}
    
    Let us examine several of the most commonly used DDF systems to understand their distributed execution models and broad design choices. We will then compare these systems with our novel approach described in Section \ref{sec:hp_ddf}.
    
        \subsubsection{Dask DDF}
        
\begin{figure}[htpb]
        \begin{center}
        \includegraphics[scale=0.5]{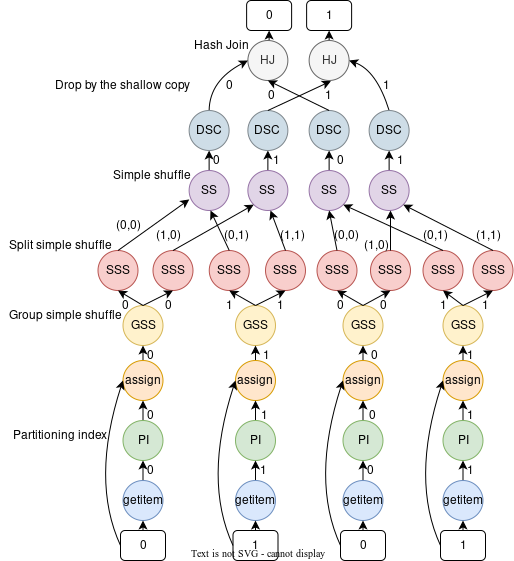}
        \end{center}
        \caption{Dask DDF \texttt{join} (2 partitions)}
        \label{fig:dask-join}
        \end{figure}

        Dask DDF is a distributed DF composed of many \pandas\ DFs partitioned along the vertical axis. Operators on Dask DDFs are decomposed into tasks which are then arranged in a DAG (Figure \ref{fig:dask-join} depicts a Join operation). Dask-Distributed Scheduler then executes these tasks on Dask-Workers. This DDF execution is a good example of AMT model. Core local operators are offloaded to \pandas{}. Communication operators (mainly \texttt{shuffle}) support point-to-point TCP message passing using \textit{Partd} disk-backed distributed object store.  
        
        \subsubsection{Ray Datasets}
        
        Ray Datasets is a DDF-like API composed of Apache Arrow tables or Python objects stored in the distributed object store. Similar to Dask, distributed operators (\textit{Transforms}) follow the AMT model. Interestingly, they support a \textit{task strategy} as well as an \textit{actor strategy}. The latter is recommended for expensive state initialization (e.g. for GPU-based tasks) to be cached. As per communication, a map-reduce style \texttt{shuffle} is used which maps tasks to partition blocks by value and then reduces tasks to merge co-partitioned blocks together. Essentially, Ray communication operators are backed by the object store. For larger data, the documentation suggests using a \textit{push-based shuffle}. 
        
        \subsubsection{Apache Spark Dataset}
        
        It is fair to say that Apache Spark is the most popular actor-based data engineering framework available today, and it has attracted a large developer community since its initial publication, \textit{Resilient Distributed Datasets (RDDs)}\cite{zaharia2012resilient}. \textit{PySpark Dataset} is a DDF-like API, and recently a \pandas{}-like DDF named \textit{Pandas on Spark} was also released. Similar to AMT, Spark decomposes operators into a collection of map-reduce tasks, after which a manager process schedules these tasks in executors allocated to the application. It uses \textit{Akka-Actors} to manage the driver (i.e. the process that submits applications), the manager, and executors. Essentially, Spark implements AMT using the actor model for map-reduce tasks. All these processes run on a Java Virtual Machine (JVM), and could face significant (de)serialization overheads when transferring data to and from Python. As an optimization, the latest versions of PySpark enable Apache Arrow columnar data format. 
        
        \subsubsection{Modin DDF}
        
        Modin \cite{petersohn2020towards} is the latest addition to the DDF domain. It introduces the concept of \textit{DF algebra} (Figure \ref{fig:cylon_modin}), where a DDF operator can be implemented as a combination of core operators. It executes on Dask \& Ray backends, which also provide the communication means for DDF. Modin distinguishes itself by attempting to mirror the \pandas{} API and follow eager execution.

\section{\cylon{} \& \cflow{}: High Performance DDFs in Dask \& Ray} \label{sec:hp_ddf}

    
Through our research, we have encountered several performance limitations while using the aforementioned DDF systems for large datasets. As discussed in Section \ref{sec:exp}, many of these DDFs show limited scalability, and we believe the limitations of the AMT model could be a major contributor to that. A centralized scheduler might create a scheduling bottleneck. Additionally, the lack of a dedicated optimized communication mechanism further compounds the issues. It is fair to assume that the optimization of communication routines is orthogonal to designing distributed computing libraries such as Dask/Ray, and re-purposing generic distributed data-sharing mechanisms for complex communication routines may lead to suboptimal performance when used in DDF implementations. 
 
In a recent publication we proposed an alternative approach for DDFs that uses BSP execution model, which we named \cylon{} \cite{widanage2020high}. It is built on top of MPI and uses MPI collective communication routines for DDF operator implementations. MPI libraries (OpenMPI, MPICH, IBM-Spectrum) have matured over the past few decades to employ various optimized distributed communication algorithms, and \cylon{} benefits heavily from these improvements. It also profits from data parallelism and implicit coalescing of local tasks by employing the BSP model. Experiments show commendable scalability with \cylon{}, fittingly differentiating it as a \textit{high performance DDF (HP-DDF)}. Even though high performance DDFs seem encouraging, having to depend on an MPI environment introduces several constraints. MPI process bootstrapping is tightly coupled to the underlying MPI implementation, e.g. OpenMPI employs PMIx. As a result, it is not possible to use MPI as a separate communication library on top of distributed computing libraries such as Dask/Ray. Usually these libraries would bootstrap their worker processes by themselves. There is no straightforward way for the MPI runtime to bind to these workers. 

We strongly believe it is worthwhile to expand on the HP-DDF concept beyond MPI-like environments. Current advancements in technology and the high demand for efficient data engineering solutions encourage this position. Our main motivation for this paper is to develop an execution environment where we could strike a balance between the scalability of BSP and the flexibility of AMT. Dask and Ray have proven track records as distributed computing libraries. So rather than building a new system from scratch, we focused on bridging the gap between BSP and these libraries. We propose a two-pronged solution to this problem. First, creating a stateful pseudo-BSP execution environment using the computing resources of the execution runtime. This lays the foundation for HP-DDF execution. The second step is using a modularized \textit{communicator} abstraction (i.e. interface that defines communication routines) that enables pluging-in optimized communication libraries. We named this project \cflow{}, as it embraces the idea of \textit{managing a workflow}.


    \subsection{Stateful Pseudo-BSP Execution Environment} \label{sec:hpddf_actor}
    
     Within this pseudo-BSP environment, executors initialize an optimized communication library and attach it to the state of the executor. The state would keep this communication context alive for the duration of an \cflow{} application. This allows \cflow{} runtime to reuse the communication context without having to reinitialize it, which could be an expensive exercise for larger parallelisms. Once the environment is set up, the executors implicitly coalesce and carry out local operations until a communication boundary is met. The state can also be used to share data between \cflow{} applications as discussed in Section \ref{sec:share_data}. 
     
\begin{figure}[htpb]
    \begin{center}
    \includegraphics[width=0.45\textwidth]{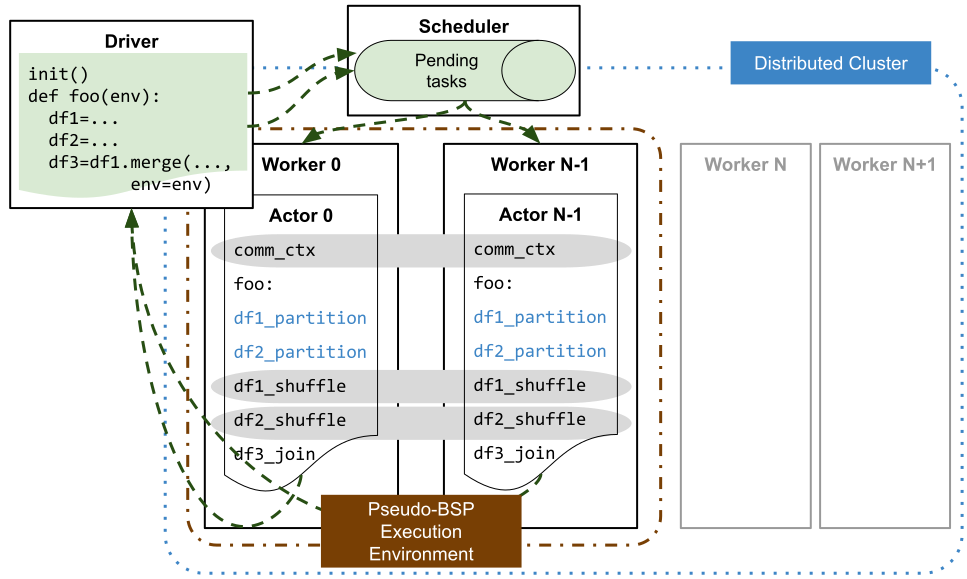}
    \end{center}
    \caption{\cflow{} Using Actors for HP-DFF}
    \label{fig:pseudo-bsp}
    \end{figure}

     This proposition of creating stateful objects matches perfectly with the actor model. Thus we leveraged the actor APIs available in Dask and Ray to implement \cflow{}-on-Dask and \cflow{}-on-Ray (Figure \ref{fig:pseudo-bsp}). An actor is a reference to a designated object (\texttt{\cylon{}Actor} class) residing in a remote worker. The driver/user code would call methods on this remote object, and during the execution of this call, \cflow{} runtime passes the communication context as an argument. Inside these methods, users can now express their data engineering applications using \cylon{} DDFs.
     
    This approach enables partitioning of the cluster resources and scheduling \textit{independent applications}. It would be a much more coarsely grained work separation, but we believe the abundance of computing units and storage in modern processor hardware, and their availability through cloud computing, could still sustain it. To the best of our knowledge, this is the first time actors are being used together with a dedicated communication library to develop HP-DDF runtimes.  This approach is semantically different from actors in Apache Spark, where they would still be executing \textit{independent tasks} in an AMT manner. Neither should it be confused with other orthogonal projects like \textit{Dask-MPI}, which is used to deploy a Dask cluster easily from within an existing MPI environment.

    
    Upon the initialization of the application, \cflow{} sends \cylon{} Actor definition (a class) to a partition of workers in the cluster based on the required parallelism. Workers then initialize these as an actor instance (remote object). At the same time, the actor instances initialize communication channels between each other, which is the entry point for creating \cylon{} DDFs (i.e. \texttt{\cylon{}\_env}). Instantiating an \texttt{\cylon{}\_env} could be an expensive operation, especially with large parallelism, as it opens up P2P communication channels between the remote objects. 
    
    The \cylon{} actor class exposes three main endpoints. 
    \begin{enumerate}
        \item \texttt{start\_executable}: Allows users to submit an executable class that would be instantiated inside the actor instance. 
        \item \texttt{execute\_\cylon{}}: Execute functions of the executable that accepts an \texttt{\cylon{}\_env} object and produces a Future.
        \item \texttt{run\_\cylon{}}: Execute a lambda function that accepts an \texttt{\cylon{}\_env} object and produces a Future.
    \end{enumerate}
    
    The following is an example code which creates two \cylon{} DFs using Parquet files and performs a join (\texttt{merge}) on them. 
\begin{lstlisting}[language=python]
def foo(env:(*\cylon*)Env=None):
    df1 = read_parquet(..., env=env)  
    df2 = read_parquet(..., env=env) 
    write_parquet(df1.merge(df2, ..., env=env), ..., env=env)

init()
wait((*\cylon*)Executor(parallelism=4).run_(*\cylon*)(foo))
\end{lstlisting}

        \subsubsection{Spawning Dask Actors}
        
        Dask does not have a separate API endpoint to reserve a set of workers for an application. Consequentially, \cflow{} uses the \texttt{Distributed.Client} API to collect a list of all available workers. It then uses the \texttt{Client.map} API endpoint with a chosen list of workers (based on the parallelism) to spawn the actor remote objects. Dask actor remote objects open up a direct communication channel to the driver, which they would use to transfer the results back. This avoids an extra network hop through the scheduler and achieves lower latency. 
        
        \subsubsection{Spawning Ray Actors}
        
        Ray provides a \textit{Placement Groups} API that enables reserving groups of resources across multiple nodes (known as gang-scheduling). \cflow{} creates a placement group with the required parallelism and submits the \cylon{} Actor definition to it. In Ray documentation \cite{ray:online}, communicating actors such as this are called out-of-band communication.

    \subsection{Modularized Communicator}
    
    Once the pseudo-BSP environment is set up, \cylon{} HP-DDF communication routines can pass messages amongst the executors. However, we would still not be able to reuse the MPI communications due to the limitations we discussed previously. 
    To address this, we had to look for alternative communication libraries which could allow us to implement \cylon{} communication routines outside of MPI without compromising its scalability \& performance. We achieved this by modularizing \cylon{} communicator interface and adding abstract implementations of DDF communication routines as discussed in Section \ref{sec:ddf}. This allowed us to conveniently integrate Gloo and UCX/UCC libraries as alternatives to MPI. Communicator performance experiments in Section \ref{sec:mulit_comm} demonstrate that these libraries perform as good as if not better than MPI on the same hardware.   
    
        \subsubsection{Gloo \cite{gloo:online}}
        
        \textit{Gloo} is a collective communications library managed by Meta Inc. incubator \cite{gloo:online} predominantly aimed at machine learning applications. PyTorch uses this for distributed all-reduce operations. It currently supports TCP, UV, and ibverbs transports. Gloo communication runtime can be initialized using an MPI Communicator or an NFS/Redis key-value store (P2P message passing is not affected). Within MPI environments \cylon{} uses the former, but for the purposes of \cflow{} it uses the latter. As an incubator project, Gloo lacks a comprehensive algorithm implementation, yet our experiments confirmed that it scales admirably. We have extended the Gloo project to suit \cylon{} communication interface.
        
        \subsubsection{Unified Communication X (UCX) \cite{shamis2015ucx}}
        
        UCX is a collection of libraries and interfaces that provides an efficient and convenient way to construct widely used HPC protocols on high-speed networks, including MPI tag matching, Remote Memory Access (RMA) operations, etc. Unlike MPI runtimes, UCX communication workers are not bound to a process bootstrapping mechanism. As such, it is being used by many frameworks, including Apache Spark and RAPIDS (Dask-CuDF). It provides primitive P2P communication operations. Unified Collective Communication (UCC) is a collective communication operation API built on top of UCX which is still being developed. Similar to MPI, UCC implements multiple communication algorithms for collective communications. Based on our experiments, UCX+UCC performance is on par with or better than OpenMPI. \cflow{} would use Redis key-value store to instantiate communication channels between \cylon{} actors. 

    \subsection{Sharing Results With Downstream Applications} \label{sec:share_data}
        
    As discussed in Section \ref{sec:hpddf_actor}, this approach allows partitioning of the cluster resources and scheduling of individual applications. These applications may contain data dependencies, for example, multiple data preprocessing applications feeding data into a distributed deep learning application. However this typically produces DDFs, and it would not be practical to collect intermediate results to the driver program. We propose an \cflow{} data store (i.e. \texttt{\cylon{}\_store}) abstraction to retain these results. In the following example, \texttt{data\_df} and \texttt{aux\_data\_df} will be executed in parallel on two resource partitions, and \texttt{main} function would continue to execute the deep learning model.
    \begin{lstlisting}
def process_aux_data(env:(*\cylon*)Env=None, store:(*\cylon*)Store=None):
    aux_data_df = ...
    store.put("aux_data", aux_data_df, env=env)
    
def main(env:(*\cylon*)Env=None, store:(*\cylon*)Store=None):
    data_df = ...
    aux_data_df = store.get("aux_data", timeout=..., env=env) 
    df = data_df.merge(aux_data_df, ...)
    
    x_train = torch.from_numpy(df.to_numpy()).to(device)
    model = Model(...)
    ...
    
init()
(*\cylon*)Executor(parallelism=4).run_(*\cylon*)(process_aux_data)
wait((*\cylon*)Executor(parallelism=4).run_(*\cylon*)(main))
\end{lstlisting}
    \texttt{\cylon{}\_store} could be backed by an NFS or distributed object store (ex: Ray's Object Store). This feature is currently being developed under \cflow{}, and is mentioned here only for completeness. In instances where applications choose different parallelism values, the store object may be required to carry out a \texttt{repartition} routine.

    \subsection{\cflow{} Features}
    
    The proposed actor-based solution \cflow{} provides several benefits compared to traditional MPI-like (BSP) environments as well as distributed computing environments.
    
        \subsubsection{Scalability}
        
        Experiments show that \cflow{}-on-Dask and \cflow{}-on-Ray offer better operator scalability on the same hardware compared to Dask DDF \& Ray Datasets, which employ AMT model (Section \ref{sec:exp}). It also surpasses Spark Datasets, which uses a conventional actor model. \cflow{} provides data engineering users a high performance \& scalable DF alternative to their existing applications with minimum changes to execution environments. 

        \subsubsection{Application-Level Parallelism}
        
        Partitioning resources within a distributed computing cluster enables parallel scheduling of multiple \cflow{} tasks. These would have much more coarsely grained parallelism compared to a typical task composed of a DDF operator. A future improvement we are planning to introduce is an execution plan optimizer that splits the DAG of a DF application into separate sub-applications (e.g. coalesce an entire branch). These sub-applications can then be individually scheduled in the cluster. Outputs (which are already partitioned) could be stored in a distributed object store to be used by subsequent sub-programs. We are potentially looking at large binary outputs which can be readily stored as objects rather than using the object store for internal communication routines. This \textit{application-level parallelism} could also enable multi-tenant job submission. 
        
        \subsubsection{Interactive Programming Environment}
        
        Petersohn et al \cite{petersohn2020towards} observed that an interactive programming environment is key for exploratory data analytics. R \& Python being interpreted languages suits very well with this experience. One major drawback of \cylon{} is that it cannot run distributed computations on a notebook (e.g. Jupyter). \cflow{} readily resolves this problem by enabling users to acquire a local/remote resource (managed by Dask/Ray) and submit \cylon{} programs to it interactively. 
                
        \subsubsection{High Performance Everywhere} 
        
        The concept of \cflow{} is not limited to distributed computing libraries, but also extends to larger computing environments such as supercomputers. We are currently developing an \cflow{} extension for leadership class supercomputers. Our end goal is to enable high performance scalable data engineering everywhere, from a personal laptop to exascale supercomputers.

\section{Experiments}\label{sec:exp} 

The following experiments were carried out on a 15-node Intel\textsuperscript{\textregistered} Xeon\textsuperscript{\textregistered} Platinum 8160 cluster. Each node is comprised of 48 hardware cores on 2 sockets, 255GB RAM, SSD storage, and are connected via Infiniband with 40Gbps bandwidth. The software used were Python v3.8; Pandas v1.4; \cylon\ (GCC v9.4, OpenMPI v4.1, \& Apache Arrow v5.0); Dask v2022.8; Ray v1.12; Modin v0.13; Apache Spark v3.3. Uniformly random distributed data was used with two \texttt{int64} columns, $10^9$ rows ($\sim$16GB) in column-major format (Fortran order). Data uses a cardinality (i.e. \% of unique keys in the data) of 90\%, which constitutes a worst-case scenario for key-based operators. The scripts to run these experiments are available in Github \cite{cylon_exp}. Out of the operator patterns discussed in our previous work \cite{perera2022high}, we have only chosen \texttt{join}, \texttt{groupby}, and \texttt{sort} operators. These cover some of the most complex routines from the point of view of DDF operator design. Only operator timings have been considered (without data loading time). Input data will either be loaded from the driver to the workers or loaded as Parquet files from the workers themselves (Dask \& Apache Spark discourage the former). Data is then repartitioned based on parallelism and cached. 

We admit that in real applications, operator performance alone may not portray a comprehensive idea of the overall performance of a data engineering framework. But we believe it is reasonably sufficient for the purpose of proposing an alternative approach for execution. Dask DDFs, Ray Datasets, Spark Datasets, and Modin DDFs are only used here as baselines. We tried our best to refer to publicly available documentation, user guides and forums while carrying out these tests to get the optimal configurations.

    \begin{figure}[htpb]
    \centering
    \includegraphics[width=0.42\textwidth]{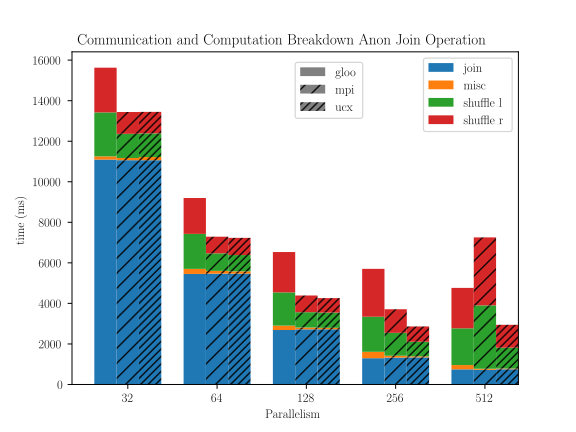}
    \caption{Communication \& Computation Breakdown of \cylon{} Join Operation (1B rows)}
    \label{fig:comm-comp}
    \end{figure}

    \subsection{Communication \& Computation}\label{sec:comm_comp}
    
    Out of the 3 operators considered, \texttt{join}s have the most communication overhead, as it is a binary operator (2 input DFs). We investigated how the communication and computation time varies based on the parallelism. Even at the smallest parallelism (32), there is a significant communication overhead (Gloo 27\%, MPI 17\%, UCX 17\%), and as the parallelism increases, it dominates the wall time (Gloo 76\%, MPI 86\%, UCX 69\%). Unfortunately, we did not have enough expertise in the Spark, Dask, or Ray DDF code base to run a similar micro-benchmark. But even while using libraries specialized for message passing, \cylon{} encounters significant communication overhead. 
        
    \subsection{OpenMPI vs. Gloo vs. UCX/UCC} \label{sec:mulit_comm}
        
    \begin{figure}[htbp]
    \begin{center}
    \includegraphics[width=0.42\textwidth]{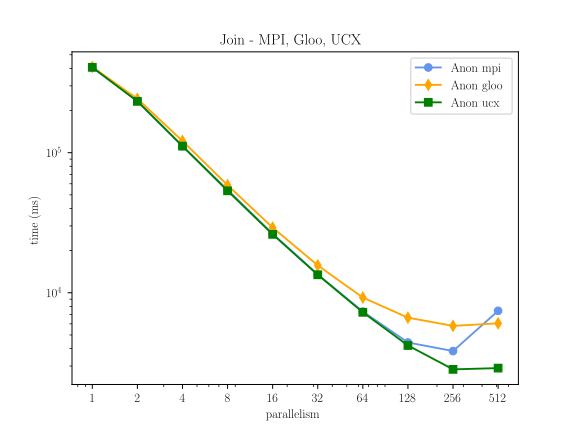}
    \caption{OpenMPI, Gloo, vs. UCX/UCC (1B rows, Log-Log) - Processes spawned by \texttt{mpirun}}
    \label{fig:mulit_comm}
    \end{center}
    \end{figure}

    In this experiment, we test the scalability of \cylon{} communicator implementations (for \texttt{join} operation). As discussed in Section \ref{sec:hp_ddf}, we would not be able to use MPI implementations inside distributed computing libraries. Figure \ref{fig:mulit_comm} confirms that our alternative choices of Gloo and UCX/UCC show equivalent performance and scalability. In fact, UCX/UCC outperforms OpenMPI in higher parallelisms. We have seen this trend in other operator benchmarks as well. 

    \subsection{\cflow{}-on-Dask \& \cflow{}-on-Ray} \label{sec:cflow_dask}
    
    \begin{figure*}[htbp]
    \begin{tabular}{ccc}
    \includegraphics[width=0.33\textwidth]{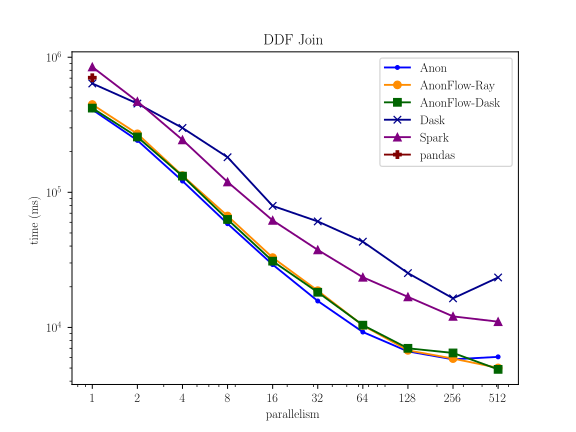} & 
    \includegraphics[width=0.33\textwidth]{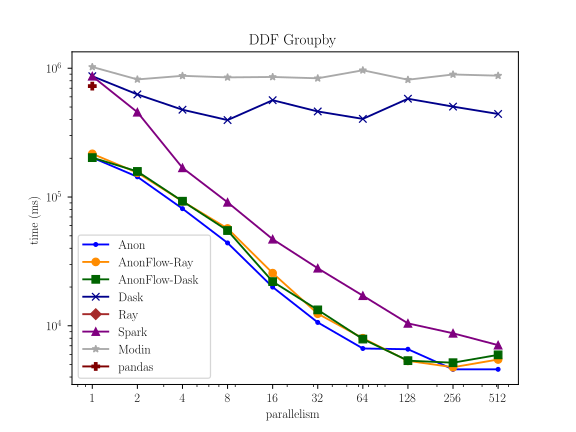} &
    \includegraphics[width=0.33\textwidth]{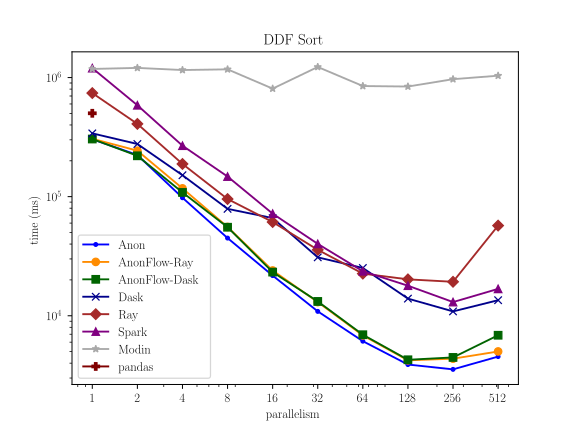} \\
    \includegraphics[width=0.33\textwidth]{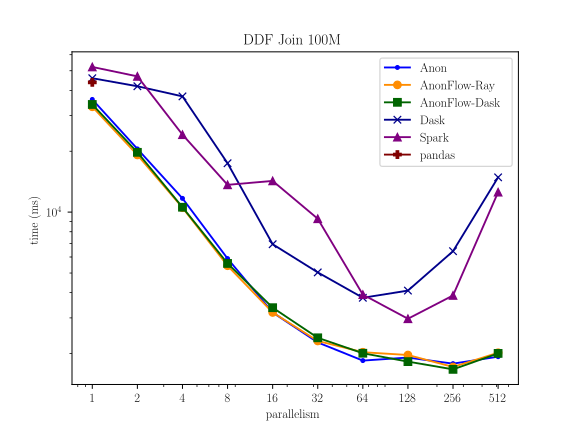} & 
    \includegraphics[width=0.33\textwidth]{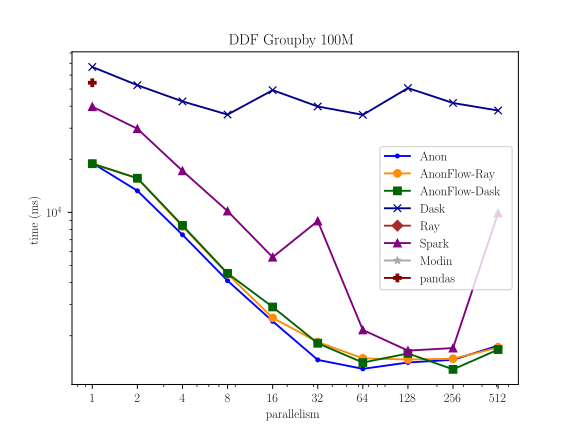} &
    \includegraphics[width=0.33\textwidth]{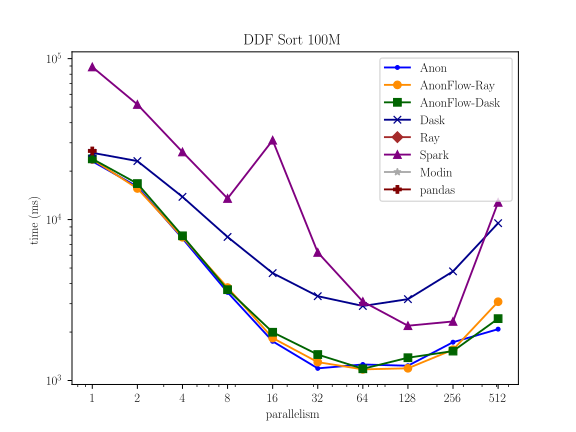} 
    \end{tabular} 
    \caption{Strong Scaling of DDF Operators (Log-Log), Top: 1B rows, Bottom: 100M rows}
    \label{fig:sperf}
    \end{figure*}
   
    In this experiment we showcase the performance on the proposed HP-DDF approach for distributed computing libraries (Dask \& Ray) against their own DDF implementations (Dask DDF \& Ray Datasets). Unfortunately we encountered several challenges with Ray Datasets. It only supports unary operators currently, therefore we could not test \texttt{join}s. Moreover, Ray \texttt{groupby} did not complete within 3 hours, and \texttt{sort} was showing presentable results. We have also included Apache Spark, since the proposed approach leverages actor model. We enabled Apache Arrow in PySpark feature because it would be more comparable.  We also added Modin DDFs to the mix. Unfortunately, it only supports \texttt{broadcast joins} which performs poorly on two similar sized DFs. We could only get Modin to run on Ray backend with our datasets, and it would default to \pandas{} for \texttt{sort}. \pandas{} serial performance is also added as a baseline comparison.  
    
    Looking at the 1 billion rows strong scaling timings in Figure \ref{fig:sperf}, we observe that \cylon{}, \cylon-on-Dask, \& \cylon-on-Ray are nearly indistinguishable (using Gloo communication). Thus it is evident that the proposed \cflow{} actor approach on top of Dask/Ray does not add any unexpected overheads to vanilla \cylon{} HP-DDF performance. Dask \& Spark Datasets show commendable scalability for \texttt{join} and \texttt{sort}, however former \texttt{groupby} displays very limited scalability. We investigated Dask \& Spark further by performing a 100 million row test case (bottom row of Figure \ref{fig:sperf}) which constitutes a communication-bound operation. Under these circumstances, both systems diverge significantly at higher parallelisms, indicating limitations in their communication implementations. We also noticed a consistent anomaly in Spark timings for 8-32 parallelism. We hope to further investigate this with the help of the Spark community. \cflow{} also shows decreasing scalability with much smoother gradients and displays better communication performance. These findings reinforce our suggestion to use a pseudo-BSP environment that employs a modular communicator. In fact, our preliminary tests suggested that using UCX/UCC communicator could potentially improve the performance further in the same setup (Section \ref{sec:mulit_comm}). 

    At 512 parallelism, on average \cflow{} performs $142\times, 123\times$, and  $118\times$ better than \pandas{} serial performance for \texttt{join}, \texttt{groupby}, and \texttt{sort} respectively. We also observe that the serial performance of \cflow{} outperforms others consistently, which could be directly related to \cylon{}'s C++ implementation and the use of Apache Arrow format. At every parallelism, \cflow{} distributed performance is $2-4\times$ higher than Dask/Spark consistently. These results confirm the efficacy of the proposed approach.
    
    \subsection{Pipeline of Operators}
    
    \begin{figure}[htbp]
    \begin{center}
    \includegraphics[width=0.42\textwidth]{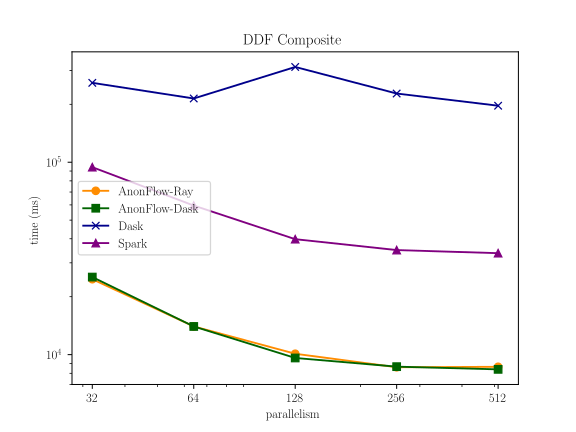}
    \end{center}
    \caption{Pipeline of Operators (1B rows, Log-Log)}
    \label{fig:composite}
    \end{figure}
    
    We also tested the following pipeline on \cflow{}, Dask DDF, \& Spark Datasets,   {\scriptsize\fcolorbox{black}{white}{\texttt{join}}$\rightarrow$\fcolorbox{black}{white}{\texttt{groupby}}$\rightarrow$\fcolorbox{black}{white}{\texttt{sort}}$\rightarrow$\fcolorbox{black}{white}{\texttt{add\_scalar}}}. As depicted in Figure \ref{fig:composite}, the gains of \cflow{} become more pronounced in composite use cases. Average speed-up over Dask DDFs ranges from $10-24\times$, while for Spark Datasets it is $3-5\times$. As mentioned in Section \ref{sec:hp_ddf}, \cylon{} execution coalesces all local operators that are in-between communication routines in the pipeline, and we believe this is a major reason for this gain.

\section{Limitations \& Future Work}

From our findings in Section \ref{sec:hp_ddf}, the idea of using BSP execution environments is a very common use case in HPC and supercomputing clusters, and the  \cflow{} concept readily fits these environments. We are currently working with Radical-Cybertools and Parsl teams to extend \cflow{} to leadership class supercomputers based on workflow management software stack. In addition, we plan to extend \cflow{} on top of pure actor libraries such as Akka. This would enable \cylon{}'s native performance on the JVM using Java Native Interface (JNI). We are currently adding these JNI bindings to \cylon{} \& \cflow{}.

In Section \ref{sec:exp} we saw significant time being spent on communication. In modern CPU hardware, we can perform computation while waiting on communication results. Since an operator consists of sub-operators arranged in a DAG, we can exploit \textit{pipeline parallelism} by overlapping communication and computation. Furthermore we can also change the granularity of a computation such that it fits into CPU caches. We have made some preliminary investigations on these ideas, and we were able to see significant performance improvements for \cylon{}. Section \ref{sec:hp_ddf} proposed an \cflow{} data store that allows sharing data with downstream applications. This work is still under active development. 

Providing fault tolerance in an MPI-like environment is quite challenging, as it operates under the assumption that the communication channels are alive throughout the application. This means providing communication-level fault tolerance would be complicated. However, we are planning to add a checkpointing mechanism that would allow a much coarser-level fault tolerance. Load imbalance (especially with skewed datasets) could starve some processes and might reduce the overall throughput. To avoid such scenarios, we are working on a sample-based repartititoning mechanism.

\section{Related Work}

In a previous publication we proposed a formal framework for designing and developing high performance data engineering frameworks that includes data structures, architectures, and program models \cite{kamburugamuve2021hptmt}. Kamburugamuve et al proposed a similar big data toolkit named \textit{Twister2} \cite{kamburugamuve2020twister2}, which is based on Java. There the authors observed that using a BSP-like environment for data processing improves scalability, and they also introduced a DF-like API in Java named \textit{TSets}. However, \cylon{} being developed in C++ enables native performance of hardware and provides a more robust integration to Python and R. Being an extension built in Python, \cflow{} still manages to achieve the same performance as \cylon{}. 

In parallel to \cylon{}, Totoni et al also suggested a similar HP-DDF runtime named \textit{HiFrames} \cite{totoni2017hiframes}. They primarily attempt to compile native MPI code for DDF operators using \texttt{numba}. While there are several architectural similarities between \textit{HiFrames} and \cylon{}, the latter is the only open-source HP-DDF available at the moment. The former is still bound to MPI, hence it would be impractical to use it in distributed computing libraries like Dask/Ray. 

Horovod utilizes Ray-actors that use Gloo communication for data parallel deep learning in its \textit{Horovod-on-Ray} project \cite{Horovodo26:online}. From the outset, this has many similarities to \cflow{}-on-Ray, but the API only supports communications on tensors. \cylon{}/\cflow{} is a more generic approach that could support both DFs \& tensors. In fact, these could be complementary frameworks, where data preprocessing and deep learning are integrated together in a single pipeline. 

In addition to the DDF runtimes we discussed in this paper, we would also like to recognize some exciting new projects. Velox is a C++ vectorized database acceleration library managed by the Meta Inc. incubator \cite{pedreiravelox}. Currently it does not provide a DF abstraction, but still offers most of the operators shown in Figure \ref{fig:cylon_modin}. Photon is another C++ based vectorized query engine developed by Databricks \cite{behm2022photon} that enables native performance to the Apache Spark ecosystem. Unfortunately, it has yet to be released to the open source community. Substrait is another interesting model that attempts to produce an independent description of data compute operations \cite{substrai7:online}.

\section{Conclusion}

Scalable dataframe systems are vital for modern data engineering applications, but despite this many systems available today fail to meet the scalability expectations. In this paper, the authors present an alternative approach for scalable dataframes, \cflow{}, which attempts to bring high performance computing into distributed computing runtimes. Their proposed stateful pseudo-BSP environment and modularized communicator enable state-of-the-art scalability and performance on Dask and Ray environments, thereby \textit{supercharging them}. \cflow{} is compared against Dask and Ray's own dataframe systems as well as Apache Spark, Modin, and \pandas{}. Using \cylon{} HP-DDF C++ backend and Apache Arrow format give \cflow{} superior sequential performance to the competition. Modular communicator in \cflow{} allows swapping Gloo and UCX/UCC for DDF communications, which enables scalable distributed performance on Dask/Ray environments. In essence, \cflow{} creates a ubiquitous data engineering ecosystem that unifies both HPC and distributed computing communities. 

\bibliographystyle{IEEEtran}   
\bibliography{ref.bib}

\end{document}